# Instagrammable Data: Using Visuals to Showcase More Than Numbers on AJ Labs Instagram Page

MATHIAS-FELIPE DE-LIMA-SANTOS[1]
University of Navarra, Spain

ARWA KOOLI
University of la Manouba, Tunisia

News outlets are increasingly developing formats dedicated to capturing audience attention in social platforms. Meanwhile, the use of data-driven storytelling is becoming increasingly integrated into the ever-complex business models of news outlets, generating more impact and visibility. Previous studies have focused on studying these two effects separately. To fill this literature gap, this study identifies and analyzes the use of data journalism on social media content of AJ Labs, the team dedicated to producing data-driven and interactive stories for the Al Jazeera news network. Drawing upon a mixed-method approach, this study examines the use and characteristics of data stories on Instagram. Results suggest that there is reliance on producing visual content that covers topics such as politics and violence. In general, AJ Labs relies on the use of reproducible formats and produces its own unique data. To conclude, we suggest potential ways to improve the use of Instagram to tell data stories.

The concept of data journalism has matured among news organizations over the years. Collecting, filtering, and analyzing data sets are becoming gradually integrated into the ever-complex business models of news outlets around the world. Thus, data journalism, once conceived as "fluid and mutable" (Hermida & Young, 2019, p. 45), has shifted from the niche and innovation spheres "to that of delivery" (Bradshaw, 2021, p. 393). That is to say, getting people to actually use it.

---

Mathias-Felipe de-Lima-Santos: mathias.felipe@unifesp.br
Arwa Kooli: arwa.kooli@gmail.com

[1] The authors disclosed receipt of the following financial support for the research, authorship, and/or publication of this article: Al Jazeera Media Institute's fellowship and the project JOLT—Harnessing Data and Technology for Journalism (H2020—MSCA-ITN-2017; Grant No. 765140).

During the COVID-19 pandemic, many news outlets were able to leverage their use of data and technology to report on the global death toll and infection rates. However, in an age with an abundance of content and increasing demands upon audience attention, news outlets have to work symbiotically with technology platforms to capture this attention, resulting in novel news formats such as Facebook Instant Articles and YouTube videos. Thus, we can contend that the attention economy has affected news production and consumption (Myllylahti, 2020), leading practitioners to incorporate these platforms as part of the news routine.

Social media platforms have given opportunities to media outlets to reach broader audiences and engage with them more dynamically (Bucher & Helmond, 2018). These platforms have also brought challenges and revealed the weaknesses in the business models of the media companies, requiring changes in the logic of news production and consumption (Sacco & Bossio, 2017).

As far as we know, no previous research has investigated the use of data journalism in social media platforms, especially on Instagram. The purpose of this study is to investigate how a particular social media platform (Instagram) has been incorporated in the data journalism workflow by Al Jazeera (AJ), an important news network in the Arab region. The network is known for producing content directed toward citizens of the Global South, where there is a substantial audience for mobile news (Zayani, 2021). Additionally, AJ provides services beyond the Arabic language, relying on English to serve wider audiences. Although this is not the only social media platform where AJ Labs—Al Jazeera's data and interactive team—publishes content (there is also a page on Facebook and on Twitter, neither of which is covered in this study), the importance of Instagram is perceived by its growth exponentially in recent years. Furthemore, media companies are still figuring out how they can benefit from Instagram (Vázquez-Herrero, Direito-Rebollal, & López-García, 2019).

Scholars have focused on studying data journalism from a Western perspective (Knight, 2015; Stalph, 2018; Zamith, 2019), but these studies tend to be limited to a few countries. Conversely, the Arab region has received little attention despite the important data-driven endeavors that have continuously developed in the region, such as InfoTimes (Egypt) and Al Jazeera (Qatar). Previous studies have indicated that journalism in the Arab region is emotive, which challenges the development of evidence-based reporting (Bebawi, 2019). Furthermore, the access to information is marginal in the Arab region because of the lack of an open data culture and freedom of information laws—except for Lebanon, Morocco, Sudan, Tunisia, Yemen, and Jordan (Lewis & Nashmi, 2019; Salimi, 2019). Despite being an outlier, Al Jazeera suffers from similar challenges to its peers in the Arab region, such as the lack of open data (Lewis & Nashmi, 2019). This makes AJ an interesting case study.

Drawing upon a literature review of data journalism and social media platforms, this study aims to understand how Instagram can be more than just an afterthought for data journalism. Following a mixed-method approach that combines content analysis and semistructured interviews with members of AJ Labs and their collaborators, we propose to answer the following research questions:

*RQ1: What are the features of AJ Labs' data stories on Instagram?*

*RQ2: To what extent does AJ Labs abide by the norms and routines of data journalism put forward by Western literature to produce data-driven content for Instagram? In what manner is AJ Labs able to develop different (i.e., its own) practices and routines?*

*RQ3: How does AJ Labs use Instagram's affordances to visualize data?*

*RQ4: How can Instagram be used to drive data narratives and become more than a tool to promote Web-based data journalism content?*

This project aims to understand how AJ Labs is creating data stories for Instagram, one of the most popular contemporary platforms for visual content (Vázquez-Herrero et al., 2019). Additionally, this study explores ways to attract audiences and make data stories more accessible and understandable to the public. The article concludes by examining several implications of this approach and proposing better ways to take advantage of Instagram's technological affordances, that is, for "understanding and analyzing social media interfaces and the relations between technology and its users" (Bucher & Helmond, 2018, p. 235).

## Theoretical Background

### *Platforms: A Symbiotic Relationship With News Media*

Since the advent of the Internet and its proliferation through mobile, journalism has increasingly become entangled with the digital ecosystem. This means that digital technologies have lowered the barriers for the public to access information, thereby facilitating distribution and enabling more users to consume media content (Rifkin, 2014). Over time, social networking sites (SNSs), defined as:

> Web-based services that allow individuals to (1) construct a public or semipublic profile within a bounded system, (2) articulate a list of other users with whom they share connections, and (3) view and traverse their list of connections and those made by others within the system. (boyd & Ellison, 2007, p. 211).

These SNSs have evolved from purely personal use to commercial use, exploiting users' data as part of their business models (Evens & Van Damme, 2016). Thus, SNSs turned into social media platforms, which gave rise to the age of platformization (Helmond, 2015).

Social media platforms have also allowed different news consumption patterns to take place across this ecosystem, while "traditionalists" might prefer online media institutions; "social news users" opt for social and interactive media experience (Geers, 2020, p. 349). This emerging media ecology has changed users' roles from passive consumers of information to active "producers" (Bruns, 2018) who contribute to news stories. This forced a change in the pattern of production across the industry, which moved away from a one-to-many model toward a many-to-many approach (Belair-Gagnon et al., 2019). Media outlets no longer exclusively produce news for the masses, and this shift requires new forms of communicating and engaging with the public.

News organizations' presence on these platforms is important not only to promote content and increase the traffic to news portals (Sacco & Bossio, 2017) but also as a new form to reach, communicate with, interact with, and engage with different audiences (Holton & Lewis, 2011). People are searching for short up-to-date news stories presented in innovative formats in these settings (Vázquez-Herrero et al., 2019).

In parallel, users have embraced mobile as the primary way to access the Internet in recent years (Dunaway, Searles, Sui, & Paul, 2018). Thus, "each platform is adopted by somewhat different communities for different purposes, developing community-specific norms and conventional practices" (Yarchi, Baden, & Kligler-Vilenchik, 2021, p. 103). In general, these affordances can be understood as material artifacts such as media technologies that allow people to interact with social media platforms (Bucher & Helmond, 2018).

To cope with this new ecosystem, news media outlets had to adapt their business models to these new parameters while simultaneously dealing with steadily shrinking advertisement revenue that was lost to these platforms and the big tech companies behind them (Myllylahti, 2018). Concurrently, news media organizations had to bring their content to social media platforms, which have a "monopoly over their audiences and their audiences' attention" (Myllylahti, 2020, p. 569). News organizations were forced to adopt a mobile-first approach (Westlund, 2013) while modifying the reporting formats to bring new journalistic outputs to social media platforms.

Despite these users looking for the same type of content, each platform has its own audience. Social media platforms, such as Twitter and Facebook, have older user bases compared with Instagram, which has a younger audience (in general, 30 and younger; Li et al., 2021). For some news organizations, Instagram is a tool to interact with sources, share stories and content, or bring behind-the-scenes information to the public, while others use it to enhance the brand's visibility and interaction with young audiences (Vázquez-Herrero et al., 2019). This can be partly explained by the reliance on media platforms and mobile devices by the majority of young users to consume news content (Westlund, 2013). In this sense, news outlets embraced more interactive, mobile-friendly, and visually enhanced content as a way to gain audience attention.

Therefore, these platforms require different know-how on the part of media outlets to better understand and engage with the public (Larrondo, Fernandes, & Agirreazkuenaga, 2017). For example, Al Jazeera launched AJ+ in 2014 to produce content that primarily targeted the "young 'mobile-first' generation of intensively connected users whose communication and social habits revolve around mobile technology" (Zayani, 2021, p. 32). In doing so, AJ+ was built to embed mobile features and attributes in the content design itself, aiming to "build audiences directly on social media and across various platforms" (Zayani, 2021, p. 32). The initiative embraced specific mobile-friendly formats and adopted a multiplatform approach to gain audience attention through short videos and GIFs. Similarly, there are a few pieces that relied on "visual explainers that use statistics and data visualization to present issues in a dynamic way" (Zayani, 2021, p. 32). Thus, Al Jazeera brings an important discussion about the role of data storytelling on other platforms.

***Data Journalism: A Fluid and Mutable Practice***

The reliance on data in journalism is not new. Investigative journalism relied on the analyses of computer records and government databases to bring checks and balances to established centers of power, ensuring the difficult equilibrium that forms the core of democracy (Coddington, 2015). Thus, "statistics have long been a staple of daily news" (Nguyen & Lugo-Ocando, 2016, p. 4). In fact, the evolution of computer-assisted reporting to data journalism is much related to the availability of data and the low-cost technologies, which brought a new level of scrutiny to journalists (Coddington, 2015). Data journalism brought us the promise of more transparency (Lewis & Westlund, 2015), interactivity (Appelgren, 2018), and greater diversity of content. Since the natural evolution of data journalism is fluid and mutable, it has resulted in a lack of a common definition of the practice (Loosen, Reimer, & De Silva-Schmidt, 2020). However, data stories that contain "a substantial element of data or visualization" (Knight, 2015, p. 59) or present quantitative elements along with visual representation (Zamith, 2019) are usually defined as data journalism.

The growing body of the data journalism literature focuses on award-winning journalistic pieces, which neither represent the day-to-day data journalism nor the daily content consumed by the public. Aiming to expand this discussion, Stalph (2018) developed an analytic framework to examine daily data stories of four elite news organizations: *Zeit Online*, *Spiegel Online*, *The Guardian,* and *Neue Zürcher Zeitung*. For this, he defined four dimensions of daily data journalism that include formal characteristics, data visualizations, data sources, as well as form and content. Stalph (2018) showed that daily data journalism predominantly covers political topics, and more than three-quarters of all cases contain both discrete text and visualization components. Additionally, these visualizations were restricted to two items on average, with every third being a bar chart. Stalph (2018) also cast a spotlight on issues of preprocessed public data as "every fourth visualization explicitly indicates governmental bodies as the data source and every fourth story uses at least one governmental data source" (p. 1348).

A more recent study brought this discussion to the United States. Zamith (2019) argued that general data journalism still has a long way to develop, as the data stories consist of simple visualizations and analyses. In his perspective, the technological opportunities of online journalism are still restricted, as most articles he surveyed did not include interactive elements or included quite limited interactivity. In line with Stalph (2018), Zamith (2019) found that the data sources came from governmental bodies; hence, most of the stories also covered policy and political topics.

Similarly, Knight (2015) conducted a content analysis of the United Kingdom's national papers. Her findings led to the conclusion that data stories covering social issues were the most common. Moreover, there was a strong prevalence of visual impact through infographics, charts, and maps. In the UK national papers, the data sources included governments, research institutes, and pan-national organizations. Knight (2015) pointed out that there was an evolution of data journalism from print to digital, which offers more options such as multimedia content, interactivity, searchability, and sharing.

Scholarly literature has not shown favorable results concerning open data principles and policies (Bebawi, 2019). In general, journalists face problems with information provided in formats that are not machine-readable, such as printed documents or PDF format, which demands considerable time and effort to use sophisticated recognition capabilities to extract data from them. These advanced computational methods are not always common among journalists from the Arab region (Fahmy & Attia, 2021). Similarly, prior studies also pointed out that elite outlets did little original data collection (Knight, 2015; Stalph, 2018), hampering the ability to produce data stories.

However, little is known about the use of data journalism on social media platforms, which is where the digital audiences are located. Retaining individual users' attention has proved to be a challenge for news outlets, and many have failed to overcome it (Myllylahti, 2020). For instance, "engaging movement or eye-catching colors" to shed light on the most relevant fact or surprising data point are suggested by practitioners (Segger, 2018). Additionally, annotations and labeling are important tools to help people understand the visual information they are presented with (Segger, 2018). On the other hand, there is a need to adjust some of the workflows, as every platform has different features that impact how users interact and engage with content. For example, Deutsche Welle experimented with augmented reality (AR) and face filters through Instagram stories to present data in novel formats to the public (Späth, Lopez, & Giefer, 2021). Similarly, La Nación (Argentina) relied on these Instagram filters to produce environmental data stories (de-Lima-Santos, 2022). However, most practitioners are still using social media platforms to promote their work. With this in mind, this contribution seeks to expand the literature by presenting the case study of Al Jazeera. It is, therefore, an initial approach to the subject.

## Methodology

The emergence of Instagram disrupted the social media scene more than one decade ago. Since its creation in 2010, many users have migrated to the photo-sharing platform from other social media platforms, making it one of the primary modes of communication around the globe. Instagram has attracted not only users but also brands to produce content and market products and services (Bozzi, 2020; Salleh, 2014). Media companies slowly adopted this workflow, offering new channels to communicate with audiences and capture their attention (Mellado & Alfaro, 2020; Ohlsson & Facht, 2017). As Instagram's uses grew, it has gradually become a subject of study for several authors (Highfield & Leaver, 2015; Omena, Rabello, & Mintz, 2020).

In this study, we chose a mixed-method design that included both quantitative and qualitative components to understand how AJ Labs is making data journalism more appealing. We also sought to find out which approach fits best on Instagram. First, we collected all posts and metadata from AJ Labs' Instagram account (@aj_labs) on April 30, 2021, using CrowdTangle, a public insights tool owned and operated by Facebook since 2016. Previous studies used this tool to trace the dissemination of COVID-19 and 5G rumors during the first wave of the coronavirus pandemic (Bruns, Harrington, & Hurcombe, 2020). In doing so, the authors highlighted some constraints, such as limited access to fully public spaces on the wider Facebook or Instagram platforms as well as some incomplete metrics (Bruns et al., 2020). Despite these shortcomings, CrowdTangle proves to be an important source of social media data for the scholarly community. Thus, we relied on CrowdTangle's application programming interface (API) connection to crawl the timeline of @aj_labs, which resulted in 107 posts and their metadata, including photos, videos, and albums. In the present context of this study, we decided to remove 18 posts that do not represent data-driven content, such as the celebration of awards or recognition for the team members or photos without a data component. The final data set for analysis was composed of 89 entries.

The data were analyzed to determine the posts' content and identify commonalities and differences in the data stories. Our analysis consists of five dimensions: basic features, data source, data visualization, audiences, and form and content. The measures were drawn from existing frameworks that aimed at classifying data stories in news portals (Stalph, 2018; Tandoc & Oh, 2017; Young, Hermida, & Fulda, 2018) or Instagram content (Vázquez-Herrero et al., 2019). Additionally, these measures were refined based on an initial qualitative assessment of the content. To examine each of these dimensions, we established and operationalized relevant categories for each. In the appendix (https://doi.org/10.5281/zenodo.6496214), a table presents an overview of the dimensions, categories, and variables used in this study. Both authors classified the posts. In the event of a discrepancy, both authors reached a consensus. Our interrater reliability was 0.97, exceeding the generally accepted minimum bond (Zamith, 2019).

Next, we triangulated the data that arose from the Instagram data stories with semistructured, in-depth interviews with practitioners working at AJ Labs or collaborating with the team. This study also draws upon interviews with pioneering members of the AJ Labs team who have left the network and whose insights and testimonies shed light on the formative phase of the deployment of data stories for Instagram. In total, five practitioners were interviewed in English via Zoom and Google Meet. This included members who are still part of the AJ Labs team and others who previously worked at AJ Labs, including editors, producers, journalists, and designers. For anonymity and standardization purposes, we did not provide further information about the participants, such as age or gender, although their roles are explicitly described in their quotes. These interviews were conducted between February and May 2021 and covered five domains: history of AJ Labs, working dynamics, data storytelling practices, audience engagement and motivation, and production of data journalism for social media, particularly for Instagram. On average, each interview lasted 50 minutes.

To analyze these interviews, we used thematic analysis using a deductive approach (i.e., the themes were developed from existing concepts or ideas that emerged from quantitative analysis (Braun & Clarke, 2006). To assist in this step, we performed it in NVivo, a proprietary software tool for qualitative data analysis. This step provided insights into the types of stories posted by AJ Labs and the motivations behind them, which will be discussed in the following section.

## Findings

### *Basic Features*

AJ Labs has built a team that is interested in experimenting with new formats and platforms while exploring and testing the latest technologies. "Most of the time, it's a day-to-day job, but there was like 10 or 20% that we would bring in new tools or ideas" (designer 1). From these experiments, AJ Labs decided to adopt a digital content strategy: "We cover one subject as simple as this, but it makes its way onto all of our platforms" (AJ Labs' data editor). This allowed the team to "think of data journalism as a process across platforms" (AJ Labs' data editor). It also allows these practitioners to learn how to embrace the ubiquity of social media and incorporate its use on data-driven content. Thus, "we think of data journalism as the entire pipeline" (AJ Labs' data editor).

However, there are some significant differences in the formal characteristics of data-driven content produced for Instagram, as shown in Figure 1. Contrary to digital and print data stories (Knight, 2015; Stalph, 2018; Zamith, 2019), we did not find authorship in all Instagram posts. This shows that AJ Labs opts for a distinct strategy in the data-driven social media content that omits traditional author bylines. In part, this is because the audience's attention is limited:

> On Instagram, the point is that you are going to appear on somebody's feed or someone's going to view your story. It's the same rule as putting up a billboard. You have somebody's attention for three or four seconds. How much information can I put into 1000 × 1000 pixels and make sure that someone's going to immediately get it? So, for Instagram, we tend to simplify a lot of our information because we don't have an audience for that long (designer 2).

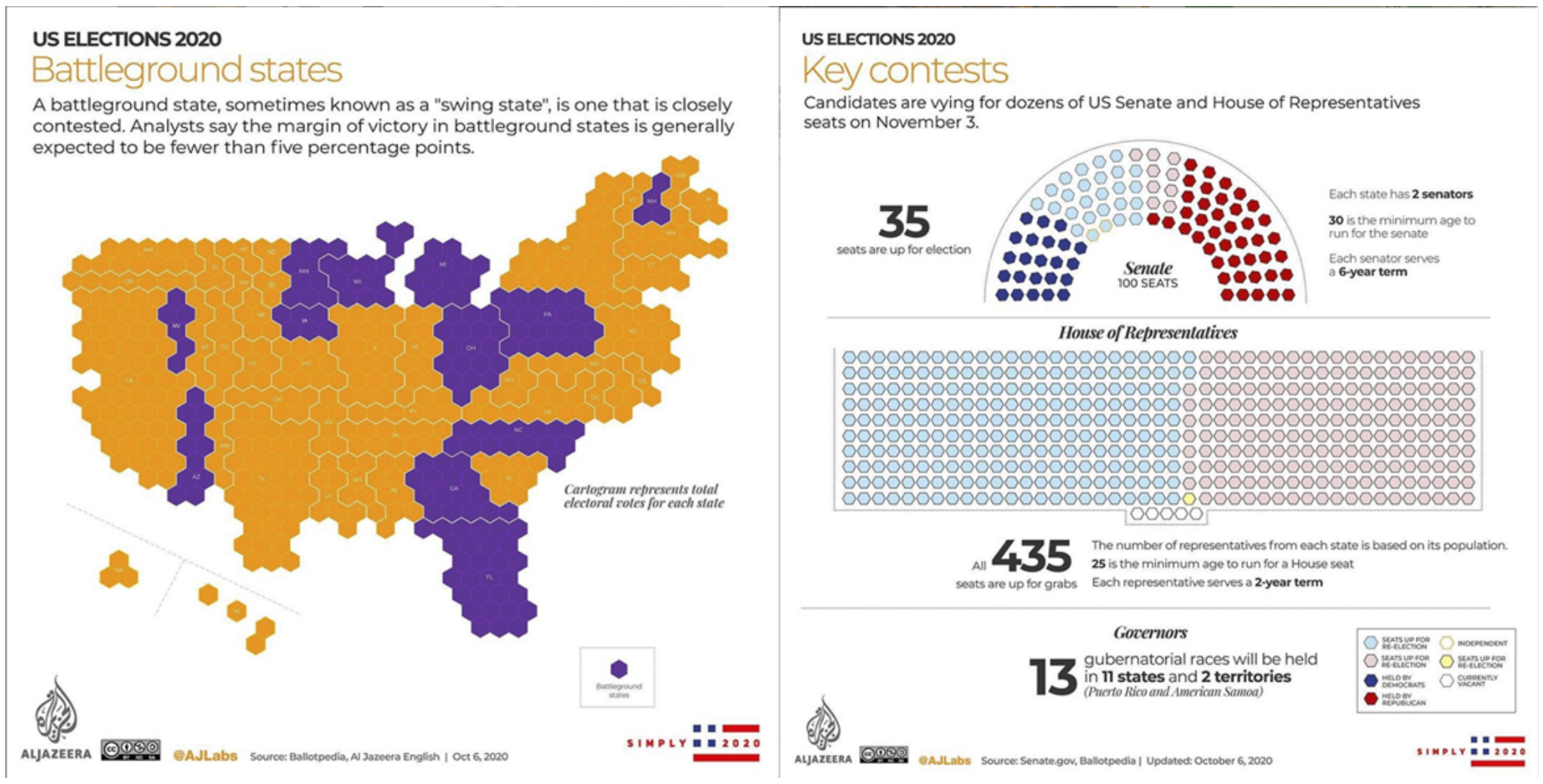

***Figure 1. Example of Instagram posts.***

This can also be explained by the data journalism workflow adopted in Al Jazeera. Our interviewees mentioned two reasons for working with editorial staff to produce data-driven content. For example, as shared by one producer, there is no official workflow for the production of posts. "If I find a new idea, I just go to AJ Labs' office and we pitch it" (producer). This convenience has facilitated cooperation between AJ Labs and other desks. On one hand, the AJ Labs team size maximizes the utility of its members by working with journalists. On the other hand, AJ Labs experiences regular fluctuations in the number of staff members, requiring it to cooperate with other members in the newsroom. It is essential, therefore, that this collaboration occurs. In some cases, this involves many professionals, which would use more space to include authorship than is available on Instagram.

Another important finding, as shown in Figure 2, reveals that there is no specific period in which AJ Labs produces more content for Instagram. Instead, there is a series of "peaks and valleys" that can be associated with various topics covered. For example, the posts on October 2017 were related to the Israeli-Palestinian conflict. Similarly, AJ Labs published different Instagram posts about Saudi Arabia and Iran, which are regional rivals and are engaging in a number of Middle East conflicts, such as the global oil trade. The most recent peak was inspired by a trend called #30DayMapChallenge, in which the team published some maps shedding light on topics related to the 2020 U.S. presidential election.

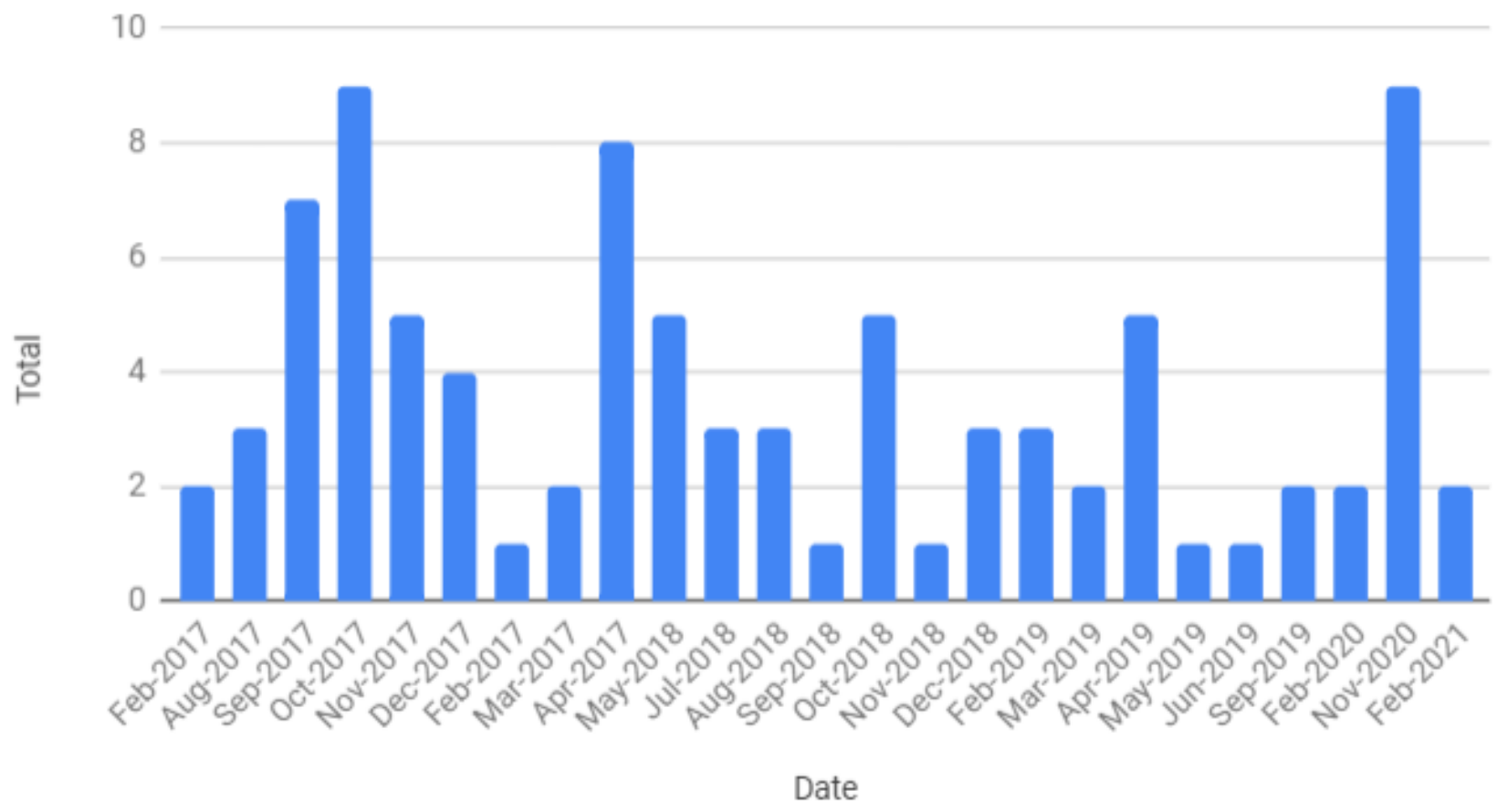

***Figure 2. The number of posts per month.***

The description has an average of 179.49 (standard deviation, $\sigma$ = 131.58) characters and about 20 words (mean, $\mu$ = 19.70; $\sigma$ = 17.96). This demonstrates that AJ Labs relies more on visual content than on the textual aspect of the stories, although there is some level of variance with an extreme value of 681 characters and 88 words. Similarly, there is low usage of hashtags in the posts ($\mu$ = .96; $\sigma$ = 1.73).

***Table 1. Type of Post Published by AJ Labs.***

| Type | Total | Percentage |
|---|---|---|
| Album | 6 | 6.74% |
| Photo | 81 | 91.01% |
| Video | 2 | 2.25% |

As Table 1 illustrates, the majority of data-driven Instagram posts were single photos (91.01%), followed by the album features, which combined more than one image in a single post. Videos, however, were limited to only two posts.

As shown in Table 2, topics related to violence are the second most popular. It is by now generally accepted that the current situation in the Arab region, with the deteriorating human rights situation associated with terrorism and illegal migration, makes this a salient theme in the media agenda. This is an important venture, as AJ Labs aims to “quantify the human side [of the events] by contextualizing them through data” (AJ Labs’ data editor). Overall, these findings are in accordance with Bebawi (2019), who describes the emotive nature of Arab journalism, which frequently focuses on the human element in its coverage. This is interesting because it shows that data journalism, which is considered to be a form of evidence-based reporting, can give voice to the people affected by public bodies’ actions and inactions (de-Lima-Santos & Mesquita, 2021a), to illustrate the data as a way to "humanize the data" (de-Lima-Santos & Mesquita, 2021b, p. 1429).

***Table 2. Main Topics Covered in Instagram Posts.***

| Topic | Total | Percentage |
|---|---|---|
| Politics | 44 | 49.44% |
| Violence | 19 | 21.35% |
| Development/Economy | 10 | 11.24% |
| Science | 5 | 5.62% |
| Social issues | 3 | 3.37% |
| Religion | 3 | 3.37% |
| Environment | 2 | 2.25% |
| Population | 1 | 1.12% |
| Health | 1 | 1.12% |
| Culture | 1 | 1.12% |

Although Instagram's affordances are broad, the use of URLs in posts is currently limited on the platform to the Instagram Stories and biographies (the small area under the username where one can share some details and include a link to a website). Even under this limitation, AJ Labs places links to data stories on half of its Instagram posts (52.81%). Although this can generate traffic, Instagram's restrictions limit the success as the URLs are shown as plain text instead of clickable links. However, click-throughs are not the main objective. AJ Labs' data editor has an explanation for this. He underlines the fact that Instagram posts should contain the entirety of the information being presented, as people do not want to leave the platform to consume information:

> [Our Instagram feed] isn't meant to attract people back to the website. It's meant to put people on the platforms to consume complete stories, which basically means whether they consume it on Instagram or they consume it on our website, they still get the essence of the story. Now, of course, the website gives us much more flexibility. In the written narrative, we can explain more details, so we do link to the Instagram article or, whenever possible, link by all of these kinds of things (data editor).

### ***Basic Features***

AJ Labs adds data sources to the majority of its content in Instagram (91.01%), which is an important component of data journalism. On average, 1.6 data sources are mentioned per post. This means that the AJ Labs team relies on more than one data source in its stories. Indeed, 36 posts contained more than one data source (40.45%). An interesting fact is that producing its own data comes in second, as shown in Table 3. A popular explanation for the reliance on multiple data sets (or building their own data sets) is that access to information is marginal in the Arab region (Lewis & Nashmi, 2019). Consequently, the origin of data is also influenced by this factor.

***Table 3. Types of Data Sources Found on the AJ Labs' Instagram Posts.***

| Type | Total | Percentage |
|---|---|---|
| Multiples (posts with data from different sources) | 36 | 40.45% |
| Own data | 13 | 14.61% |
| National government | 11 | 12.36% |
| Civic society groups/NGOs | 8 | 8.99% |
| NA | 7 | 7.87% |
| Private company | 4 | 4.49% |
| International agency | 4 | 4.49% |
| International government | 3 | 3.37% |
| Academic | 3 | 3.37% |

Among the data sources indicated in the Instagram posts, we observed that these multiple sources combine numerous institutions, including civil society organizations, nongovernmental organizations, and news agencies, as shown in Figure 3. Looking specifically at these posts with data from different sources (40.45%), we could identify that a majority of these data come from international agencies (24.47%), media sources (21.28%), and their own data sets (20.21%).

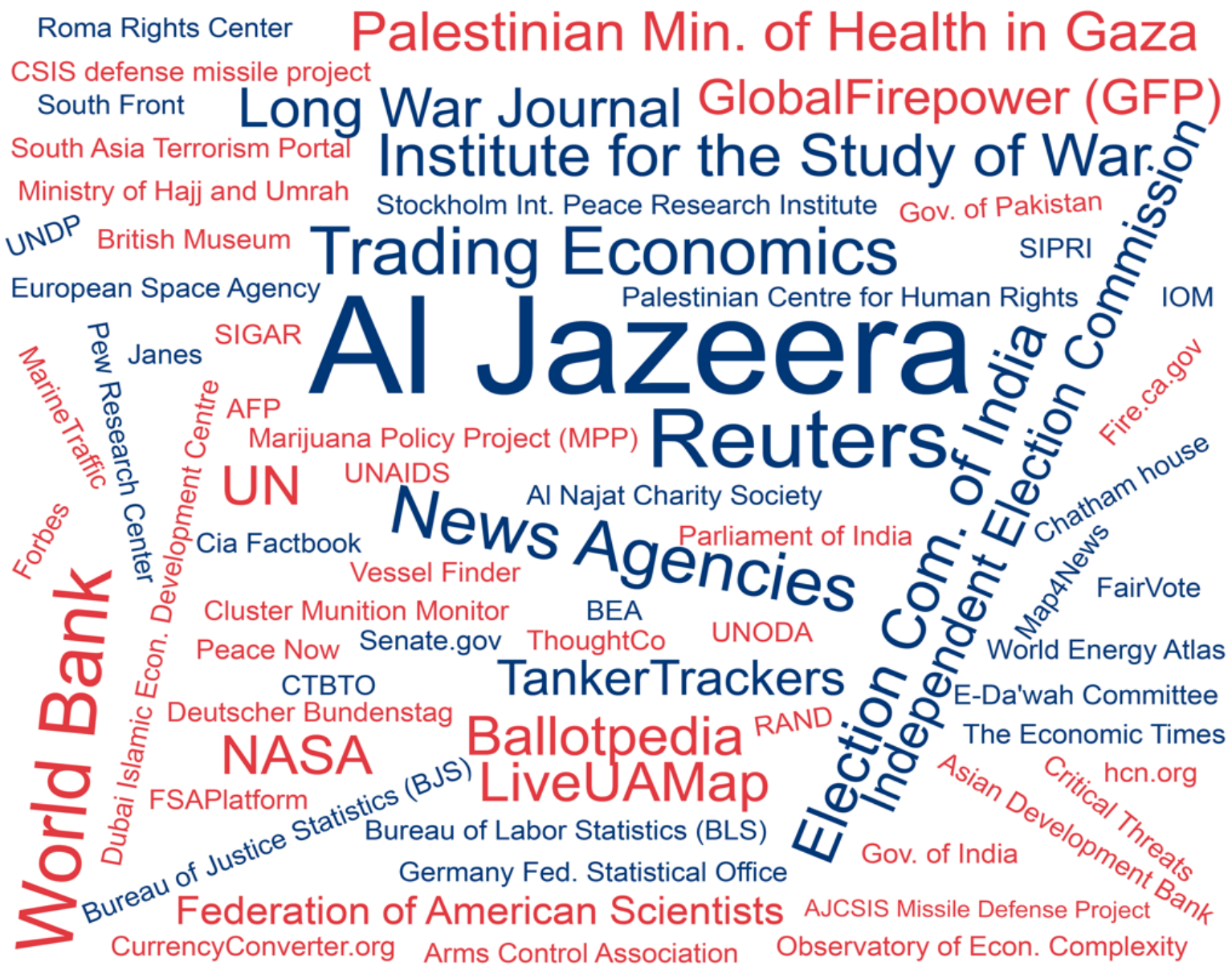

***Figure 3. Visual representation of data source indicated in the Instagram posts.***

Similarly, almost 40% of data with unique sources came from international organizations, led by the United States (20.48%), as shown in Table 4. We speculate that this might be because Al Jazeera reaches wider audiences by offering its content in English, which means that there is a greater public interest in topics from an international perspective. Al Jazeera also has a large readership base in the United States, which might explain the reliance on data sets originating from this country. Interviewees also have similar perceptions. Furthermore, "there is a wide number of U.S. data sources available, which makes easy access to these data for use in various projects" (producer). Conversely, few Arab countries appeared to be indicated as the origin of data sources, with the exceptions of Qatar and Saudi Arabia. The former is because many data sets were produced by AJ Labs, which we classified as sourced from Qatar.

***Table 4. The Origin of Data Sources.***

| Data source's country | Total | Percentage |
|---|---|---|
| International | 33 | 39.76% |
| USA | 17 | 20.48% |
| Qatar | 12 | 14.46% |
| India | 7 | 8.43% |
| Palestine | 3 | 3.61% |
| Afghanistan | 2 | 2.41% |
| Pakistan | 2 | 2.41% |
| UK | 1 | 1.20% |
| Switzerland | 1 | 1.20% |
| Sweden | 1 | 1.20% |
| Saudi Arabia | 1 | 1.20% |
| Italy | 1 | 1.20% |
| Israel | 1 | 1.20% |
| Germany | 1 | 1.20% |

The lack of open data culture and freedom of information laws in the Arab region (Lewis & Nashmi, 2019) challenged the AJ Labs team to be innovative by collecting its own unique data from unrestricted public data sets using Web scraping techniques or by processing data from open sources. For example:

> To explain to a global audience how bad the economy of Lebanon is, we produced a series of fairly simple graphics in which we quantified how bad the situation was by going to two hypermarket chains' websites, Carrefour and Spinneys, the two very large supermarkets in Lebanon. All we did was collect a huge database of all the prices of all the items over two years, so anyone can do it technically. It's very simple (AJ Labs' data editor).

### ***Data Visualization and Interaction***

To examine data visualization and interaction on Instagram data-driven posts, we analyzed AJ Lab's posts with different variables. In general, the Instagram posts relied on more visual content than text. Similarly, no interactivity is by far the most common characteristic of AJ Labs' Instagram posts (91.01%). We are aware that Instagram offers few interactivity features in the newsfeed while offering more interaction through Instagram Stories (Vázquez-Herrero et al., 2019). This includes using AR applications to augment the surroundings or interact with virtual objects in real time (de-Lima-Santos, 2022). On the other hand, there is the possibility to use the "play" feature (in videos) and "scroll right" feature (in albums), which were limited to, respectively, 2.25% and 6.74% of the posts. In other words, only 8.99% of posts include this limited interactivity.

Regarding the types of visualization, about three-fifth were infographics, which are a combination of visual elements and typographic resources. We identified infographics in 61.80% of AJ Labs posts, followed by maps (20.22%) and charts (13.48%). Interestingly, two Instagram posts made use of maps built from tiles that proportionally represent the data, also known as tiled cartograms.

Figure 4 examines the topics covered within the visualizations and demonstrates that infographics are aesthetically aligned to political topics as well as violence. Unsurprisingly, topics concerning development and economic issues are more visualized through charts. Similarly, social issues are also more represented through graphics. This is an interesting finding that can help practitioners better understand how to best visualize some types of data.

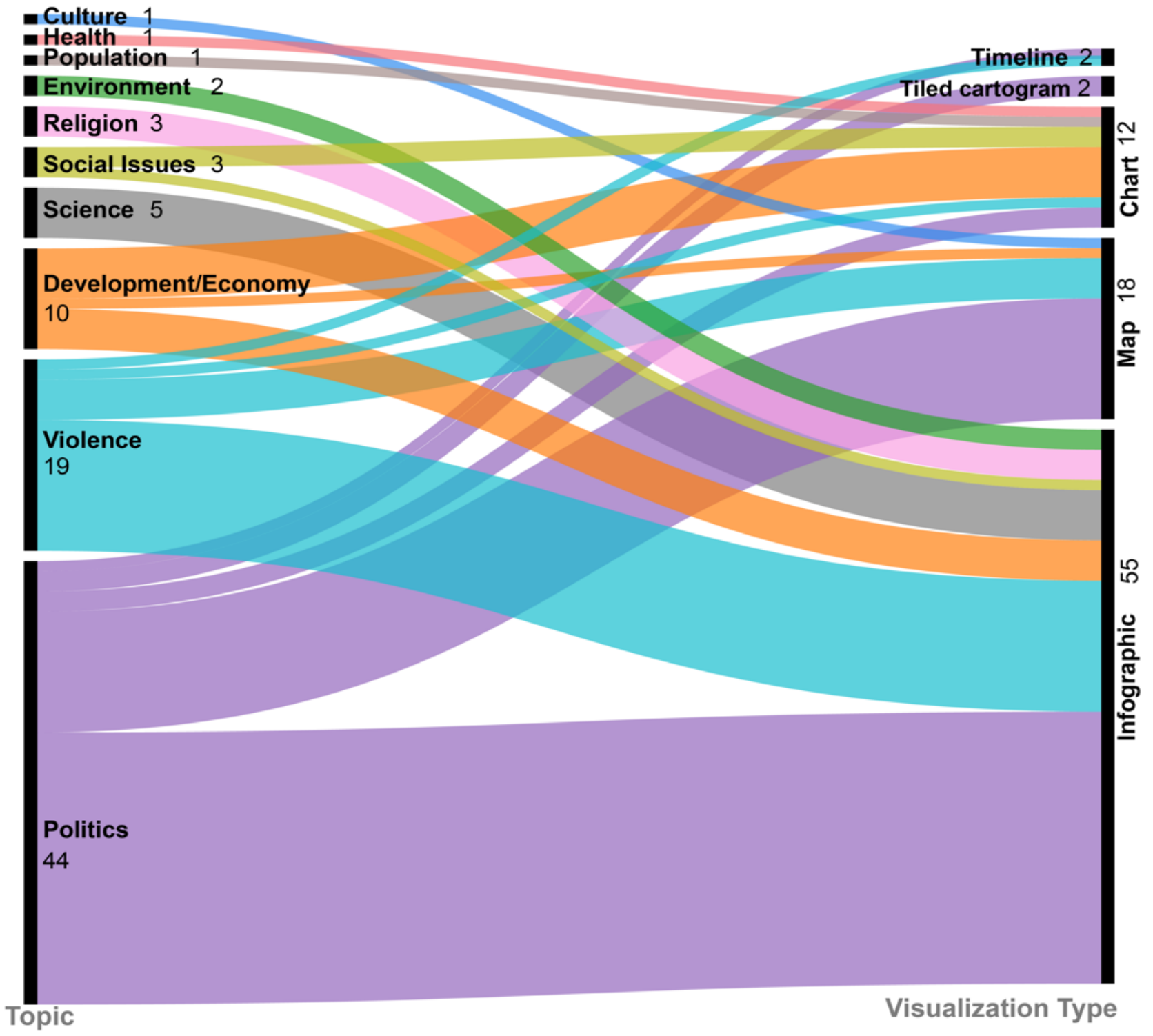

***Figure 4. Visual representation of topics and the most common visualization types for each.***

As mentioned above, the AJ Labs team is relatively small, which demands the use of third-party software to produce some visualizations. These out-of-the-box solutions offer free versions as well as paid associated services, encouraging users to upgrade for access to the premium features (de-Lima-Santos, Schapals, & Bruns, 2021). For instance, our analysis revealed that some posts relied on tools such as Carto, Flourish, and Maps4News. This is explained because the team can analyze the data, work on the story, and easily ingest it into Flourish, for example. "We use Flourish because it gives us a lot of the tools that we need and the most efficient way possible" (AJ Labs' data editor). However, the practitioners are aware of the risks posed by these platforms. For example, "it was a huge disaster for everyone who relied on [Google] Fusion Tables" (AJ Labs' data editor) when the solution was discontinued. Similarly, one of the most successful data stories produced by the team was "built on Adobe News, but it became obsolete and Adobe stopped supporting it" (designer 1). This shows how these solutions can be double-edged swords that facilitate the workflow on the one side but on the other side results in the gradual loss of data stories (de-Lima-Santos et al., 2021).

Additionally, we identified the use of reusable formats that enable the rapid transfer of information to new media (see Figure 5), as explained by the data editor:

> The graphics themselves are designed so that they could fit into the Instagram screen that is square [allowing us to reuse it on platforms]. Even the way the graphics are designed, you might notice that they are intended to be consumed in bite-sized formats. That's why they always have a title and they're always a one-line summary, and then they always have the links so that you can find more (AJ Labs' data editor).

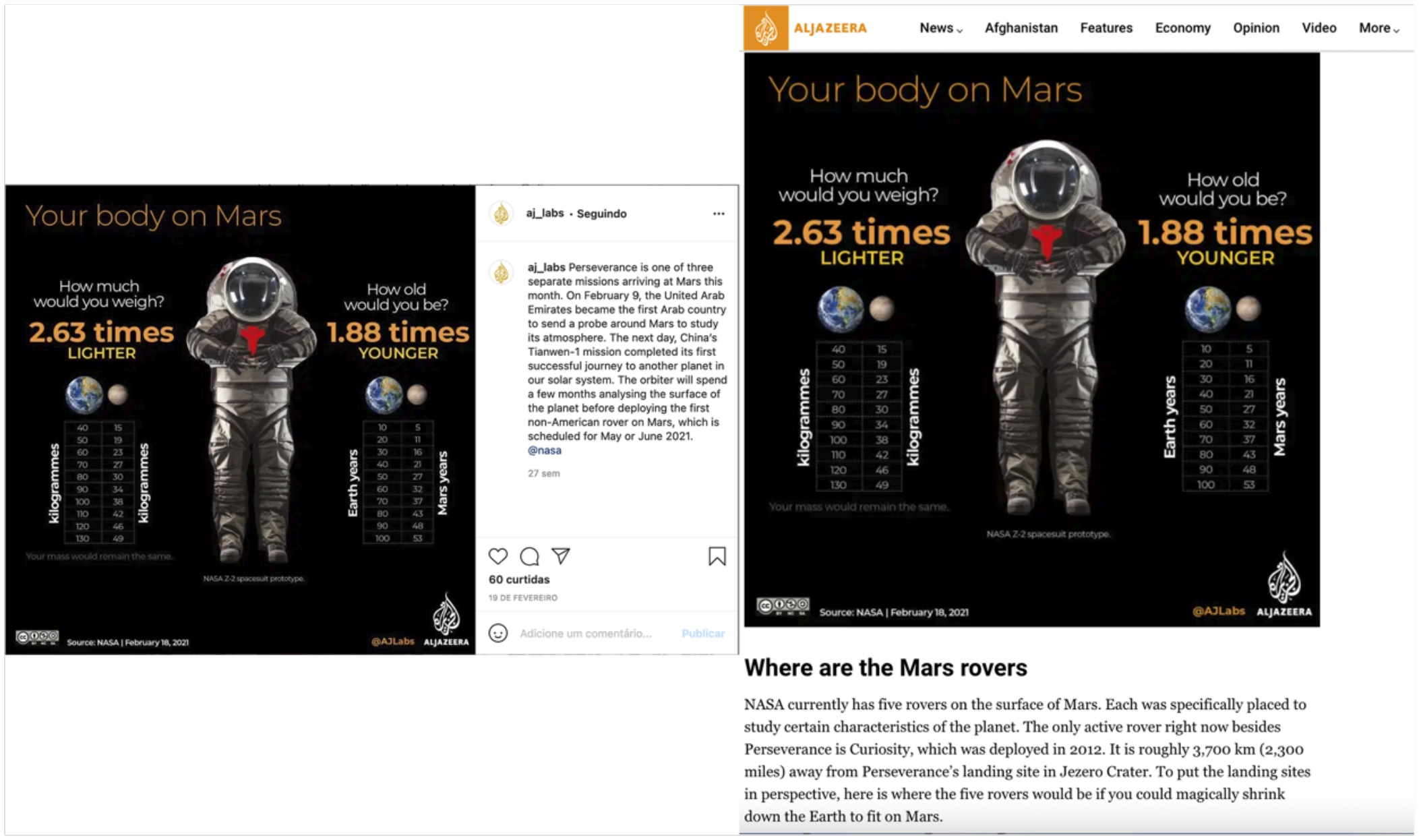

***Figure 5. Reusable formats.***

***Audiences***

Within the Instagram data, several variables allowed us to measure audience engagement to a certain extent by extracting the likes and comments in these posts, as well as the interactions. Overall, our data showed that there is no statistically significant connection between audience engagement and the posts' content. While the Instagram posts had an average of 37 likes ($\mu$ = 37.41; $\sigma$ = 40.87), the comments reached a mean of less than a post ($\mu$ = .68; $\sigma$ = 2.32). These low numbers result in a low interaction rate ($\mu$ = 38.10; $\sigma$ = 42.84), considering the more than 2,000 followers that AJ Labs had as of May 2021. However, these data also show that numbers are spread out, indicating that some are outliers.

These outlier posts provided rich information about the types of posts that generated more engagement. In particular, a story told through a collection of photos (album) describing the changes in the map of Palestine over the past 100 years had 355 likes and 21 comments. Despite the topic being very relevant in the Middle East, the format of the story might also have influenced its higher engagement. Using Instagram's affordance, the story is told through a combination of 10 photos that navigates the reader through the evolution of Palestine's history. This offers a completely different approach to telling the story and feeds the user with a full narrative, in line with the AJ Labs data editor's view:

> The graphic has to tell a story. It has to be those five graphics that should be able to give a reader who isn't ever going to come to our website because that's not what they're interested in. They want to consume on the platform, and don't try to convince them that they need to leave the platform (AJ Labs' data editor).

This suggests that the "album" and "scroll right" features should be used more frequently in Instagram posts instead of single images with no interaction.

### ***Form and Content***

Looking into the form and content of AJ Labs' Instagram posts, we were able to extract meaningful information from the text component. Following what the data editor said that the public wants to "consume on the platform" (AJ Labs' data editor), about 69.66% of the posts aim to inform the public (see Table 5). This is followed by posts that we identified as having the primary goal of generating traffic (22.47%). We related a small portion to the idea of expanding the knowledge of the users and providing extra information (6.74%).

***Table 5. Purpose of AJ Labs' Instagram Posts.***

| Purpose | Total | Percentage |
|---|---|---|
| Inform the audience | 62 | 69.66% |
| Traffic (Attract the reader to website) | 20 | 22.47% |
| Expand the knowledge | 6 | 6.74% |
| Promote (showcase project) | 1 | 1.12% |

AJ Labs relies on three main forms to tell a story in visual format (see Table 6): informative (47.19%), summary (24.72%), and explanatory (23.60%). The first encompasses posts that bring new information to the audience, such as introductory stories, while summary aims to tell the whole story in one post and explanatory brings information to clarify and explain concepts, ideas, and issues. Overall, these findings are in accordance with the purpose of these posts, as the majority aims to inform the audience.

Interestingly, we identified minor posts (3.37%) that use data to give readers future insights or projections—for example, a map of the next solar eclipses around the world.

***Table 6. Story Format Used in AJ Labs' Instagram Posts.***

| Story Format | Total | Percentage |
|---|---|---|
| Informative | 42 | 47.19% |
| Summary | 22 | 24.72% |
| Explanatory | 21 | 23.60% |
| Projection | 3 | 3.37% |
| Additional information | 1 | 1.12% |

Additionally, we examined the text captions of the posts to determine if there was any additional relevant information. Our analysis found that only one-third of the posts added content to supplement the visualization. For the rest of the posts (two-thirds), there was no relevant information. A similar pattern was identified in the use of hashtags ($\mu$ = .97; $\sigma$ = 1.73) and mentions ($\mu$ = .32; $\sigma$ = .56).

**Discussion and Conclusion**

Our study demonstrates the use of data-driven content on social media platforms. Similar to AJ+, which "led Al Jazeera to experiment with nontraditional formats and launch programs that capitalized on media convergence" (Zayani, 2021, p. 7), AJ Labs experiments with news platforms and their features. We focused on the content produced by AJ Labs for Instagram, a platform that is known for its visual content (i.e., one of the main characteristics of data journalism (Knight, 2015)).

It is generally accepted that visuals play an important role in data journalism. In Western literature, there is a strong prevalence of using visualizations in data journalism to generate visual impact, as these data-driven graphics are didactic elements from which "readers can derive information and knowledge" (Stalph, 2018, p. 1335).

To bring data stories to Instagram, AJ Labs relies on visual content, mainly infographics (RQ1). Prior works identified infographics as the most commonly used format used in projects short-listed for data journalism awards (Córdoba-Cabús & García-Borrego, 2020; Young et al., 2018). Another study has shown that infographics are also widely used in ordinary data journalism projects (Knight, 2015). In part, this is explained by the fact that the main topics covered are politics. Our result also correlates well with previous studies on data journalism wherein political topics (49.44%; see Table 2) are predominant (Stalph, 2018). For example, almost half (48.6%) of data journalism projects nominated for the Data Journalism Awards from 2013 to 2016 (a total of 179) discussed political topics (Loosen et al., 2020).

What is different is the coverage of violence (second most covered topic), reflecting the current context of the Middle East (RQ2). By quantifying the human side of data, AJ Labs relies on simple infographics to represent it, bringing some characteristics of emotive Arab journalism to its news reporting (Bebawi, 2019). Our findings, therefore, showed that data journalism in Instagram tends to use simple visualization components with discrete texts (Stalph, 2018; Zamith, 2019).

Contrary to the findings of the Western literature (RQ2), our analysis brought important distinctions concerning data sources. They are important components of data journalism (Fahmy & Attia, 2021; Knight, 2015; Zamith, 2019), but our results are not in line with previous data journalism scholarship in Western democracies, which showed that news organizations mostly rely on data from governmental bodies (Knight, 2015; Stalph, 2018; Zamith, 2019). Because the Arab world suffers from a lack of open data culture (Lewis & Nashmi, 2019), AJ Labs creates its own data using a combination of methods, such as collecting data from hypermarket chains to show how the economy has evolved in recent years. Conversely, this requires more advanced computational skills, which are not always common among journalists from the Arab region (Fahmy & Attia, 2021). This is a striking difference in Al Jazeera's technical capacity compared with its counterparts in the Arab region (Fahmy & Attia, 2021; Lewis & Nashmi, 2019). In this regard, the diversification of data methods collection is of particular importance because it provides a variety of sources and topics even in countries where access to information and data culture are given (Borges-Rey, 2016).

Interactivity is not the main characteristic of AJ Labs' Instagram posts (RQ3). Scholars have concluded that interactive elements were quite limited in daily data journalism (Stalph, 2018; Zamith, 2019), describing interaction as illusory in most cases (Appelgren, 2018). In accordance with that, AJ Labs adopted the use of reproducible formats and a digital content strategy, which allow the team to produce content for different platforms without using more resources or more time. By creating visuals that are in the formats of social media platforms, the AJ Labs team did not add extra work to its pipeline, optimizing its time and resources in favor of making its work more efficient. The use of out-of-the-box solutions also helps in this process.

To fully explore Instagram, AJ Labs should adopt some features and affordances offered by the platform (RQ4). For instance, the post that generated the most engagement used the "album" feature to tell a story in parts, helping the public interact with and absorb the information in an engaging format rather than simple images. This "scroll-right-telling" could replicate the future of "scrollytelling," commonly found in contemporary data stories. The same applies to videos, helping news outlets better connect with young generations found on this platform. Instagram Stories are some of the tools used for newsrooms to engage with the audiences in different ways and gain the public's attention (Späth et al., 2021; Vázquez-Herrero et al., 2019).

Additionally, hashtags are important features of social media platforms in general and Instagram in particular. By using hashtags, users can make their content more discoverable. Several studies have shown how news organizations are also using hashtags to index their content and reach wider audiences (Highfield & Leaver, 2015; Vázquez-Herrero et al., 2019). This reflects one of the main goals of data journalism: evolve from niche audiences to become mainstream (Stalph & Borges-Rey, 2018). With that in mind, AJ Labs and other media outlets might improve interaction and engagement with their audiences on Instagram if they use more photo albums, videos, and stories. The use of hashtags can also help these posts become indexed, thus making them more search-friendly on Instagram (Späth et al., 2021; Vázquez-Herrero et al., 2019). AJ Labs seems to play it safe by using a couple of static and simple data visualizations; however, by using different and unusual types of visuals, its Instagram page can engage a broader audience.

Furthermore, the low use of hypertextual resources might influence the reach of AJ Labs' content, as its "usage allows for a story to be linked to other publications tagged by either topic or location, as well as to connect people involved or organizations related to the content" (Vázquez-Herrero et al., 2019, p. 7). More recently, Instagram expanded its link sticker for everyone so users can share links on their Instagram Stories. This could improve AJ Labs' interactions and engagements with their audiences on Instagram and attract audiences to the news portal.

We acknowledge that this is not easy, as the frequent personnel changes in the AJ Labs team hamper the continuity of the production process. Scholars have associated the downsizing in Al Jazeera's workforce with the sharp drop in oil prices in 2016 (Zayani, 2021). It is for this reason understandable that there is no pattern associated with the number of Instagram data-driven posts produced over the past several years.

Regarding the limitations of this study, it could be argued that it relies on CrowdTangle to extract the data and metadata but cannot extract text from images. Future studies could manually conduct this conversion of text found in the images to understand the ratio between text and visualizations. As mentioned before, this is an initial approach to data journalism on social media platforms, specifically on Instagram. Future research could examine other platforms, such as Facebook and Twitter, which offer different features and affordances. In summary, our study adds to the literature by using these empirical results to understand how data journalism can permeate this medium to reach wider audiences.